\documentclass{aa}
\usepackage{epsfig,psfig,natbib,txfonts}
\usepackage{graphicx}
\newcommand{\kms}{km\,s$^{-1}$\,}

\begin{document}

%
   \title{The origin of the hot metal-poor gas in NGC~1291 \thanks
    {Based on observations obtained at the European Southern
    Observatory, Paranal, Chile (ESO program 71.B-0458A)} }
    \subtitle{Testing the hypothesis of gas dynamics as the cause of
    the gas heating}  \author{ I. P\'erez \inst{1} \and
    K. Freeman\inst{2}}  \offprints{I. P\'erez Mart\'{\i}n \\ email:
    isa@astro.rug.nl}  \institute{Kapteyn Astronomical Institute,
    University of Groningen, Postbus 800, Groningen 9700 AV,
    Netherlands \and RSAA, Mount Stromlo Observatory, Cotter road,
    Weston Creek, ACT 2611, Australia }  \date{}   \abstract{  In this
    paper we test the idea that the low-metallicity hot gas in the
    centre of NGC~1291 is heated via a dynamical process. In this
    scenario, the gas from the outer gas-rich ring loses energy
    through bar-driven shocks and falls to the centre.  Heating of the
    gas to X-ray temperatures comes from the high velocity that it
    reaches ($\approx$ 700 \kms) as it falls to the bottom of the
    potential well. This would explain why the stellar metallicity in
    the bulge region is around solar while the hot gas metallicity is
    around 0.1 solar. We carried out an observational test to check
    this hypothesis by measuring the metallicity of HII regions in the
    outer ring to check whether they matched the hot gas
    metallicity. For this purpose we obtained medium resolution long
    slit spectroscopy with FORS1 on the ESO VLT at Paranal and
    obtained the metallicities using emission line ratio
    diagnostics. The obtained metallicities are compatible with the
    bulge stellar metallicities but very different from the hot-gas
    metallicity. However, when comparing the different time-scales,
    the gas in the ring had time enough to get enriched through
    stellar processes, therefore we cannot rule out the dynamical
    mechanism as the heating process of the gas. However, the blue
    colours of the outer ring and the dust structures in the bar
    region could suggest that the origin of the X-ray hot gas is
    due to the infall of material from further out. \keywords{Galaxies: ISM --
Galaxies: structure
    -- Galaxies: kinematics and dynamics -- Galaxies: abundances --
    X-rays: galaxies } }\maketitle

\section{Introduction and hypothesis}

{\object NGC~1291} is an S(R)SB(s)0/a galaxy widely studied due to its brightness
($B_T$=9.39, de Vaucouleurs et al. 1991)\nocite{devauc91}, its large
angular size ($D_{25}$~=~9.8~arcminutes), and its distance, between
6.9 and 14~Mpc. In the blue, the morphology is characterised by a
bright inner lens, a primary bar, a small secondary bar misaligned by
$\approx$~30$^{\circ}$ with the primary bar, and a well defined outer
ring. The galaxy was first studied in great detail by~\cite{devauc},
who drew attention to the secondary bar as a new phenomenon in
barred galaxies. The large outer ring has a diameter around 8$\arcmin$
($\approx$~16~kpc). HI data show that the atomic gas is concentrated
in the optical ring, with a pronounced central hole~\citep{vandriel}. 
The derived total HI mass is 0.81~$\times$~10$^{9}$M$_{\odot}$ 
for an adopted distance of 6.9 Mpc. NGC~1291 is then relatively gas rich 
for an S0/a galaxy.\\

Recent Chandra data ~\citep{hogg,irwin} show that the hot gas is anti-correlated with the HI: the X-ray emission fills the
central hole. There are two components in the X-ray emission with evidence for decreasing 'hardness' with radius.  The Chandra data
allow the separation of the stellar emission from the gas emission,
showing that the diffuse gaseous component has a ratio $L_{x}/L_{B}=1.3 \times 10^{29}$
ergs~sec$^{-1}$~L$_{B_{\odot}}^{-1}$. This luminosity is low, but
similar to the values found for X-ray faint ellipticals and the bulge
of M31. The derived X-ray mass within a radius of 120$\arcsec$ is 5.8
$\times$ 10$^{7}$ M$_{\odot}$~\citep{irwin} which is lower than the
average mass of bright ellipticals.  They calculated a cooling time of
the gas at the centre of the galaxy of 1.2$\times$10$^{7}$ yr and less
than a Hubble time for the hot gas filling the hole.\\

The unexpected discovery from the Chandra observations is that the
derived metallicity of the X-ray emitting gas near the centre of the
galaxy is only 0.13~$\pm$~0.04~$\times$~solar. In contrast, an optical
measurement of the metallicity of the stellar bulge found it to have a
$Mg_{2}$ index of 0.24, corresponding to Fe/H~=~1.1~$\times$
solar~\citep{terlevich}. Such low X-ray gas abundances have been
observed for other early-type galaxies
~\citep{sarazin,osullivan}. Enrichment of the hot gas in early-type
galaxies is generally thought to be due to stellar processes such as
type Ia supernovae or stellar mass losses. However, the low metallicity
of the hot gas is hard to reconcile with this scenario. Therefore, a
new source of heating must be found. Dynamical heating is a
possibility. For example, we know that most of the HI is now in the
outer ring. The HI ring coincides with the stellar ring, and the sharp
edges of the ring suggest a dynamical origin, such as a resonance. The
bar will have a strong dynamical influence out to the outer Lindblad
resonance (OLR, i.e. out to the vicinity of the outer ring of NGC~1291). If one allows for the gas outside the bar to move in
non-circular orbits due to the effect of the bar, then the gas can
lose angular momentum and can dissipate large amounts of energy
through shocks and will fall in toward corotation. If one models the
potential field as an isothermal King model, and knowing that the
stellar velocity dispersion of the galaxy is about
160~\kms~\citep{dalle}, one can estimate the infall velocity. For an
isothermal, truncated distribution the central potential is
$\Phi\approx9\sigma^{2}$, the velocity is then $V=(2|\Phi|)^{1/2}$~$\approx$ 700~\kms. This velocity is enough to
heat up the gas to X-ray temperatures as it falls to the bottom of the
galaxy's potential well.\\  In this scenario, the hot metal-poor X-ray
gas near the centre of NGC~1291 comes from the gas which we already
observed to lie in the outer ring. To test this scenario, we need to
find out if the metallicity of the gas in the ring is low and therefore
compatible with the metallicity of the X-ray gas near the centre. It is worth noticing that due to the limited survey of X-ray emission in S0/a ringed galaxies, little is known about whether this low metallicity in the hot gas component is a common feature in this type of galaxies.\\

In this paper we present the observations, data reduction
(Section~\ref{data})  and abundance determination
(Section~\ref{abundance}) of a sample of HII regions in the outer ring
of NGC~1291. In Section~\ref{discussion} we present the results from
the abundance determination. The different time-scales at which the
different processes taking place occur are presented in
Section~\ref{timescales}. Finally, discussion and conclusions are
presented in Section~\ref{Conclusions}. \\

\section{Observations and data reduction}
\label{data}

\begin{figure*}
\begin{center}
\vspace{0cm}
\hspace{1cm} \hbox{\epsfxsize=8.0cm \epsfbox{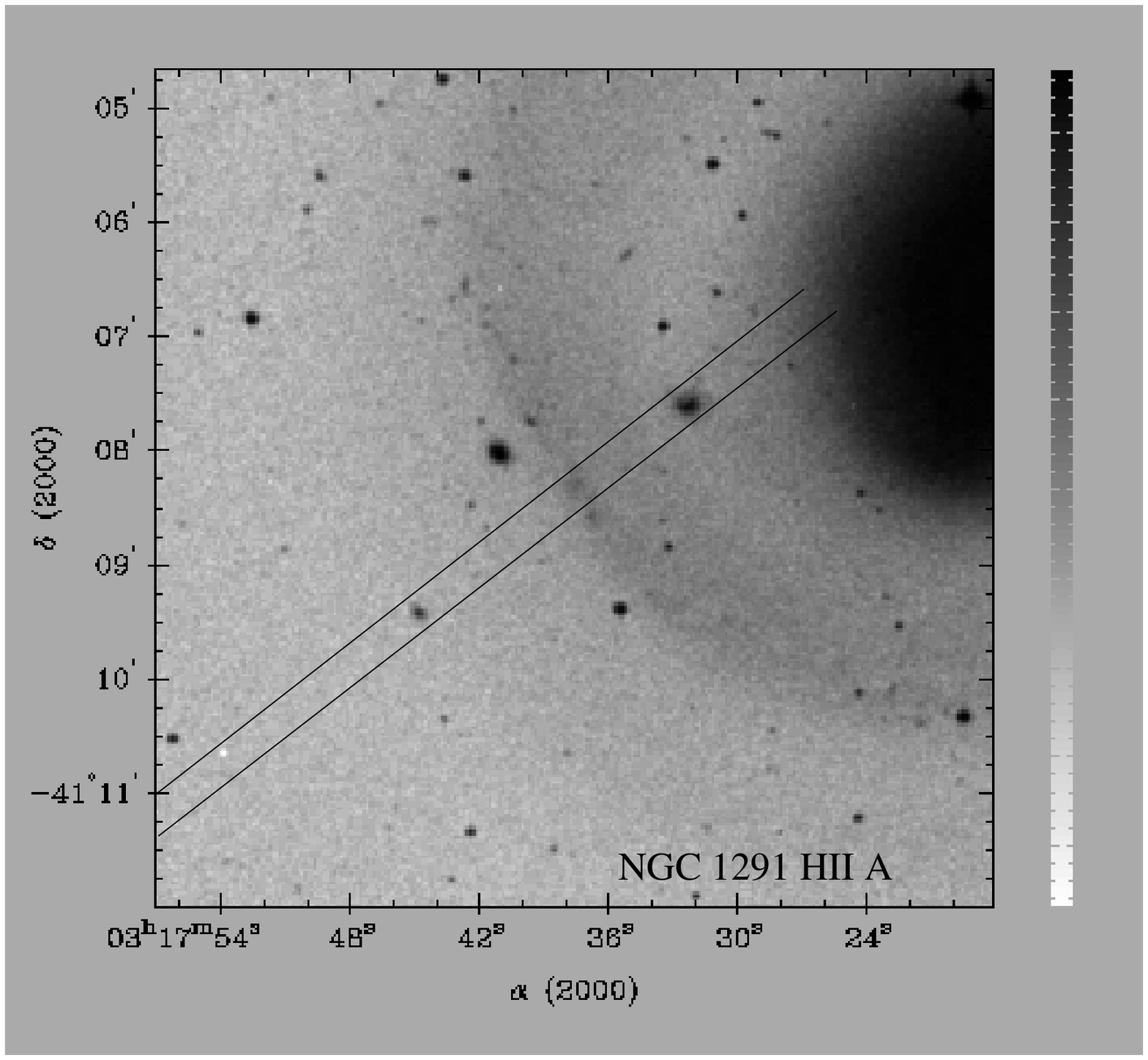}
\hspace{1cm}\epsfxsize=8.0cm \epsfbox{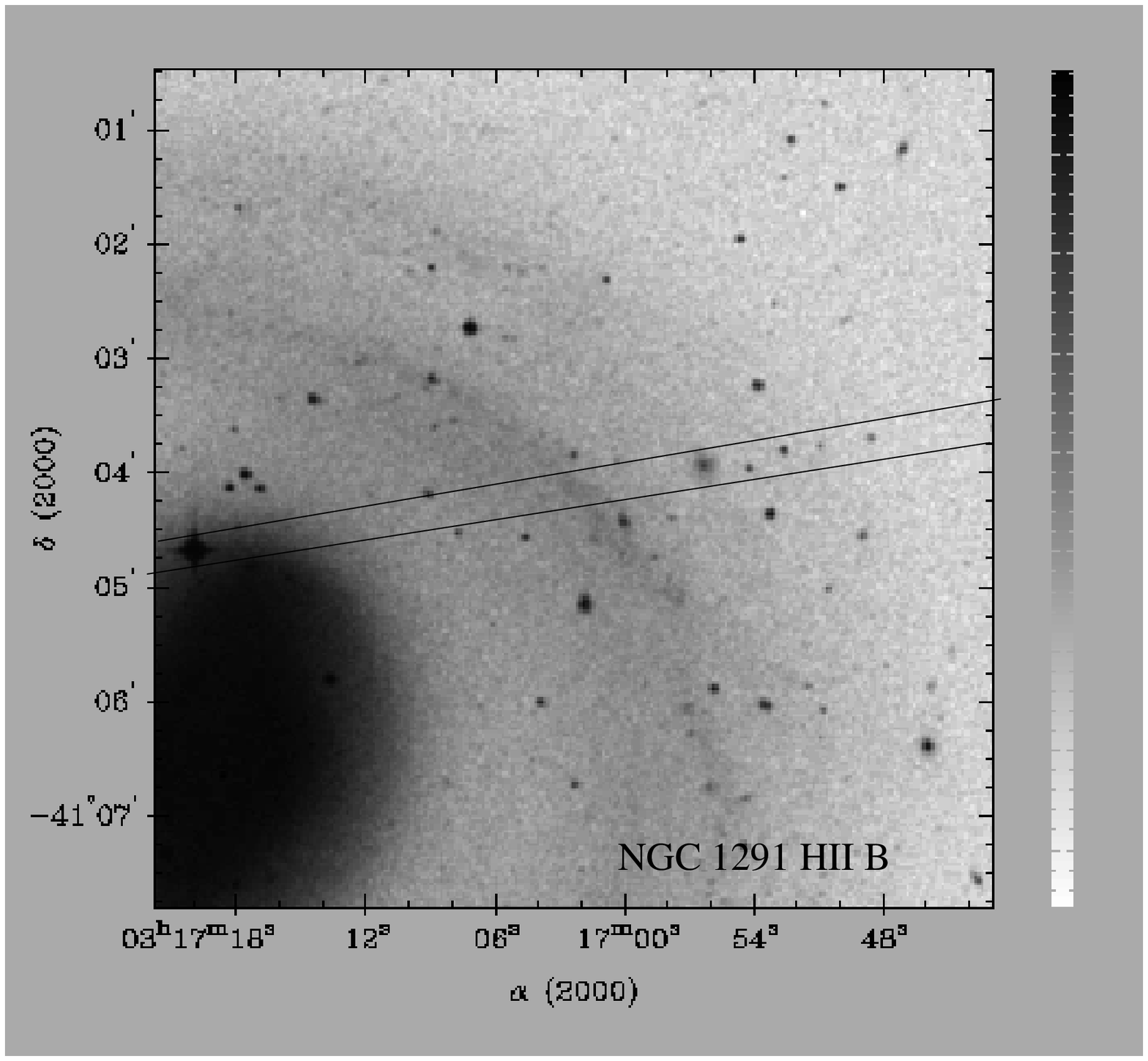}}
\vspace{1cm}
\caption{Long-slit FORS1 observations. The left panel presents the
slit position for region A on the south-east side of NGC~1291. The
right panel presents the slit position for region B on the north-west
side of NGC~1291.}
\label{fig:slits}
\end{center}
\end{figure*}

Service observations were carried out at the VLT with FORS1 at the UT1
in August 2003. Two hours were allocated for the project to obtain
long slit spectra at two different slit positions on the galaxy with
two different instrument configurations (grism  600V and grism
300V). The position of the HII regions were chosen by visual
inspection of the DSS image. As already noted by~\cite{crocker}, only
faint HII regions are found in the outer ring in contrast to the inner
region where strong H$\alpha$ emission is found. An image of the two
slit positions is shown in Fig.~\ref{fig:slits}. Observations of
1000 sec per slit position with the grism 600V were obtained. The
wavelength range covered was 4650--7100 $\AA$ with a dispersion of
49~$\AA$~mm$^{-1}$. A second set of exposures of 400 seconds were
obtained per slit position with the lower resolution 300V grism with a
wavelength range from 3300--6500 $\AA$ and a dispersion of
110~$\AA$~mm$^{-1}$. Standard calibrations were obtained together with
three spectrophotometric standards.\\

\begin{figure}
\begin{center}
\vspace{0cm}
\hspace{0cm}\psfig{figure=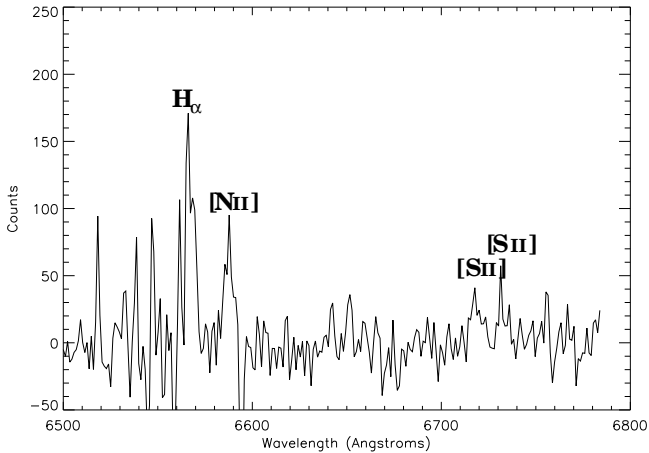,width=9.0cm}
\vspace{0cm}
\caption{Spectrum obtained for the HII region situated in the ring region
(slit position A), showing the emission lines used in the
abundance calculations.}
\label{fig:spec}
\end{center}
\end{figure}

Standard data reduction was performed using IRAF routines. All the
spectra were bias-subtracted and flat-field corrected with the standard
calibration frames provided with the service observations. The spectra
were then wavelength calibrated and geometric corrections were applied to
correct for any misalignment. The slits were centered on the outer
ring so sky subtraction could be carried out using the area outside
the galaxy. The spectra were 'flattened' using the spectrophotometric
standards acquired during the observing run. The fluxes were
obtained by fitting a Gaussian to the different emission lines using
different spatial binning (see Fig.~\ref{fig:spec}).

\section{Abundance determination}
\label{abundance}

\begin{figure}
\begin{center}
\vspace{0cm}
\hspace{0cm}\psfig{figure=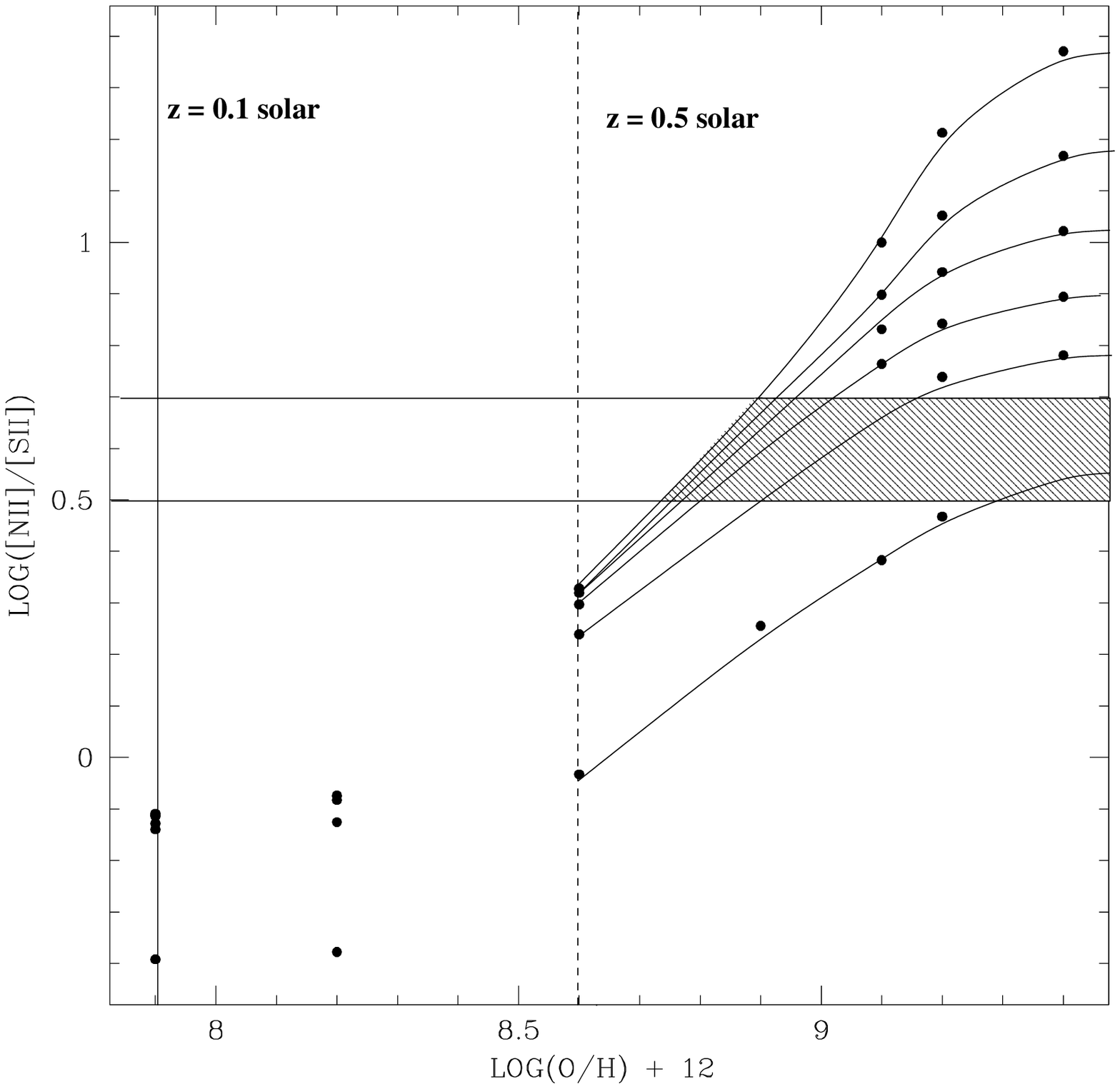,width=8.0cm}
\vspace{1cm}
\caption{Log([N{\small II}]/[S{\small II}]) {\it vs} metallicity. 
Curves are shown for ionisation parameters with
$q$=[5.6$\times$10$^{6}$, 1.0$\times$10$^{7}$, 2.0$\times$10$^{7}$,
4.0$\times$10$^{7}$, 8.0$\times$10$^{7}$] from top to bottom, and
points with metallicities 0.1, 0.2, 0.5, 1.0, 1.5, 2.0, 3.0
Z$_{\odot}$ from left to right. The shaded region shows the observed
ratios for the ring HII region with error bars included.}
\label{fig:ratio1}
\end{center}
\end{figure}

\begin{figure}
\begin{center}
\vspace{0cm}
\hspace{0cm}\psfig{figure=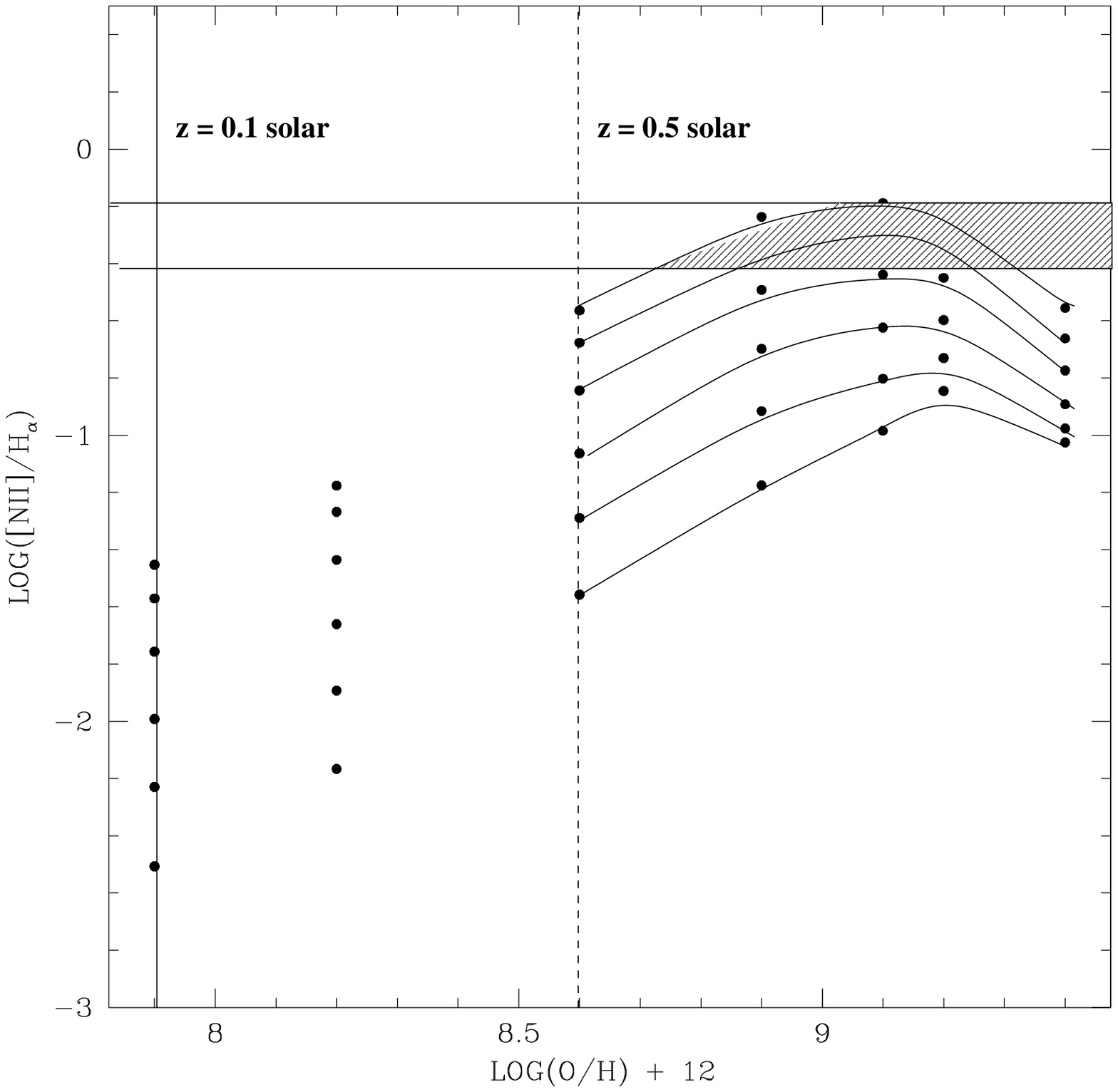,width=8.0cm}
\vspace{1cm}
\caption{Log([N{\small II}]/H$_{\alpha}$) {\it vs} metallicity.  Curves
are shown for ionisation parameter with $q$=[5.6$\times$10$^{6}$,
1.0$\times$10$^{7}$, 2.0$\times$10$^{7}$, 4.0$\times$10$^{7}$,
8.0$\times$10$^{7}$] from top to bottom, and points with metallicities
0.1, 0.2, 0.5, 1.0, 1.5, 2.0, 3.0 Z$_{\odot}$ from left to right.  The
shaded region shows the observed ratios for the ring HII region with
error bars included. Note that both diagrams give a metallicity for the
ring HII region between solar and twice solar, very different to the
0.1 solar metallicity obtained by Irwin et al.  (2002) for the hot
gas.}
\label{fig:ratio2}
\end{center}
\end{figure}

A few emission lines ([O{\small III}] 5007~\AA, H$\alpha$, [N{\small
II}] 6583~\AA, [S{\small II}] 6716 and 6731~\AA) were detected but only
in slit position A: see Fig. 1. Since H$\beta$ and [O{\small II}] were
not detected, the Balmer decrement cannot be used to estimate
extinction, and we cannot use metallicity indicators such as the
[N{\small II}]/[O{\small II}] ratio which is among the best diagnostics
for high metallicities~\citep{kewley}. Nor can we use the ionisation
parameter ({\it q}) diagnostic ratio [O{\small III}]/[O{\small II}].
We opted for using a combination of the [N{\small II}]/H$\alpha$ and
the [N{\small II}]/[S{\small II}] ratios as presented in~\citet{kewley}.
We made use of the diagrams presented in their paper to locate the
data for our H{\small II} regions on the theoretical line-ratio {\it
vs} metallicity diagrams. The advantage of using these ratios is that
the lines are close together, so no dust extinction
correction is needed.  Furthermore, the [N{\small II}]/[S{\small II}]
ratio is unaffected by the absorption lines of the underlying stellar
population. As one can see from Fig.~\ref{fig:ratio1}, at low
metallicities the ratio is quite insensitive to metallicity changes.
However, it seems a good indicator to discern whether we are in the
presence of a high or a low abundance region.  This ratio is also
dependent on the ionisation parameter, but one can get an estimate of
{\it q} using the other line ratio, [N{\small II}]/H$\alpha$. The
[N{\small II}]/H$\alpha$ ratio is very sensitive to shock excitation or
the presence of a hard ionising field (i.e. AGN). Also, the H$\alpha$
emission line will be affected by stellar absorption although, in
these regions, we expect it to be an almost negligible effect.  In
Fig.~\ref{fig:ratio1} and~\ref{fig:ratio2}, the grid computed
by Kewley, available on-line from her web page, are shown together with
the observed line ratios.  The grids correspond to HII region models
with ionising radiation derived from stellar atmospheres and stellar
evolution codes, non-LTE radiative transfer, and Salpeter IMF,
including also a consistent treatment of the dust physics~\citep{dopita}. 
Here, we use the photoionisation grids with the same parameters as 
in Kewley et al. (2002): $q$= [5.6$\times$10$^{6}$,
1.0$\times$10$^{7}$, 2.0$\times$10$^{7}$, 4.0$\times$10$^{7}$,
8.0$\times$10$^{7}$] and Z=0.5, 1.0, 1.5, 2.0 and 3.0 Z$_{\odot}$.

\section{Discussion}
\label{discussion}

\subsection {On the X-ray metallicities}
The X-ray abundances obtained by~\cite{irwin} are based on the
two-component model consisting of a MEKAL component to represent the
gaseous emission and a power law or thermal bremsstrahlung to
represent the unresolved stellar sources. The hard-component was fixed
at the value found for the sum of the resolved sources. The
temperature and metallicity of the MEKAL component were allowed to
vary.  It has been long argued that the low metallicities found in the
X-ray gas of galaxies are due to the single-temperature fit to the
spectra  and that a multi-phase fit with higher abundance represents the observed spectra~\citep{buote} equally well. However, this would account for
a difference of $\leq$~20\% in the metallicity values, which in the
case of NGC~1291 cannot explain the low metallicities found.

\subsection{Ionised gas abundances. The results}
Two HII regions were detected in one of the slit positions (position
A) while no emission lines were detected along the other slit position
(position B), see Fig.~\ref{fig:slits}. On position A, one of the HII
regions lies just outside the bulge region, near the end of the bar
(hereafter, the 'bar region') and the second one lies on the outer ring
(hereafter, the 'ring region').\\

Both regions show a high [N{\small II}]/H$_{\alpha}$ ratio indicating
already a metallicity around solar. This result also indicates that
the ionisation parameter is low, in the range 
1~$\times$~10$^{7}$~--~2~$\times$~10$^{7}$~cm~s$^{-1}$.\\

The bar region shows [O{\small III}] emission, but no [S{\small II}]
emission is observed. The log([N{\small II}]/[O{\small III}]) ratio is around
$-$0.5 which is normal for HII regions, perhaps a little low, which is
to be expected because of the lack of reddening correction, so the
combination of the ionisation parameters of
1~$\times$~10$^{7}$~--~2~$\times$~10$^{7}$~cm~s$^{-1}$ and an upper
limit log([O{\small III}]/[N{\small II}])~$<$~0.5 is consistent with a
metallicity of solar but not consistent with a metallicity 2~$\times$~solar,
using the solar abundance value from~\citet{anders}. Furthermore, the
lack of [S{\small II}] emission indicates a high value for the
([N{\small II}]/[S{\small II}]) ratio which is incompatible with a low
metallicity value. \\

[S{\small II}] emission was detected in the inner region giving a
log([N{\small II}]/[S{\small II}]) ratio of about 0.6~$\pm$~0.1, again compatible with a metallicity around solar.  A value for the
ionisation parameters of
1~$\times$~10$^{7}$~--~2~$\times$~10$^{7}$~cm~s$^{-1}$ gives a
metallicity incompatible with values higher than 1.2~$\times$~solar. \\

From the results presented in Fig.~\ref{fig:ratio1} and
Fig.~\ref{fig:ratio2}, the observed abundance of the HII regions is
clearly higher than 0.1~$\times$~solar. The difference between
1~$\times$~solar and 0.1~$\times$~solar corresponds to a factor of 10
in [N{\small II}]/H$_{\alpha}$. \\

\subsection{ Time scales}
\label{timescales}

We find that the metallicity of the ionised regions in the ring of
NGC~1291 is around solar. One could argue that the gas in the ring has
been enriched after  a few episodes of star formation from an initial
metallicity close to what is found for the X-ray emitting gas. However, the
time-scales of both processes may differ greatly.\\

\begin{figure}
\begin{center}
\vspace{0cm}
\hspace{0cm}\psfig{figure=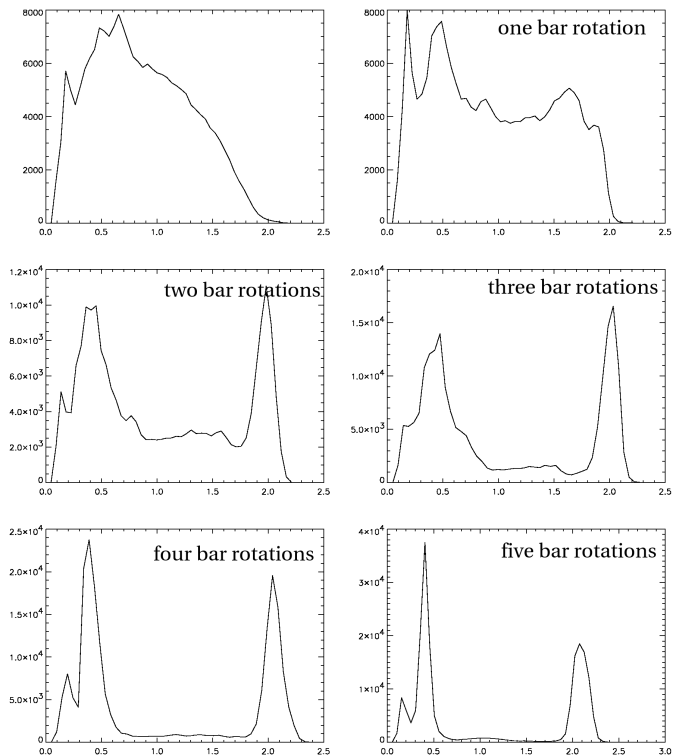,width=9.0cm}
\vspace{1cm}
\caption{Time evolution of the number of particles vs angular momentum distribution in a barred potential. The panels show the distribution up to 5 bar rotations, for a simulation with corotation at the end of the bar. The two peaks correspond to the inner and outer ring respectively.}
\label{fig:lz}
\end{center}
\end{figure}

To estimate the time scales for the gas to get from the outer disk to
the centre we ran a number of SPH simulations in rigid rotating barred
stellar potentials~\citep{perez}. The initial particle radial
distribution for the gas is a Beta function with a standard deviation
set to the scale-length of the disk and radially vanishing at a
distance 4 times the scale-length. The non-axisymmetric part of the
potential is linearly grown during three bar rotations, so the gas
flow can steadily adjust to the forcing of the bar. We observed that,
within two bar rotations after the non-axisymmetric component is
switched on (one bar rotation before the bar is fully grown), most of
the particles are re-distributed around angular momenta corresponding
to the OLR and the region inside corotation (see Fig.~\ref{fig:lz}). This reflects the
dissipation due to the shocks induced by the forcing of the bar. After
the non-axisymmetric component is fully grown the gas reaches a stable
configuration. Once the gas has settled, the angular momentum
distribution remains almost constant. We therefore estimate the
outer disk gas infall time-scale to be of the order of two bar
rotations.\\

To estimate the bar rotation speed, we assume that the corotation
radius is located at the end of the bar. The observed rotational
velocity is uncertain because the galaxy is so nearly face-on (from
isophotal analysis we derived an inclination of
6$^{\circ}$~$\pm$~2$^{\circ}$). From the Tully-Fisher (T-F) relation
\citep{sakai}, we can use the apparent magnitudes in $H$ (2MASS) and
$B$ (RC3) to estimate the rotational velocity.  We derive a rotational
velocity at the ends of the bar of $200 \pm 52$ \kms for an
adopted distance of 6.9~Mpc (the error does not include the error in
the distance). The timescale for two bar rotations is then about 300~Myr.\\

The star formation rates estimated from the H$_{\alpha}$ fluxes from
the discrete HII regions are of the order of
0.1~M$_{\odot}$~yr$^{-1}$~\citep{caldwell}. Using a simple closed-box
model the time-scale for the abundance enrichment of the ISM in the
ring from Z~$\approx$~0.1~Z$_{\odot}$ to solar metallicities is longer 
than 1 Gyr.

The total optical and near-IR colours ~\citep[2MASS]{vandriel} of
NGC~1291 are typical of early-type galaxies. However, peculiar dust
morphology is seen in the lens and the bar region. In the same region,
wispy filaments are seen in the ionised gas \citep{crocker}. The outer
ring shows bluer colours and also signs of spirality. The gas
compression in spiral shocks gives rise to star formation, which would
explain the blue colours of the outer ring. The gas content, blue
colors and spiral structure of the outer ring might point toward a
minor merger event or gas accretion, and therefore an external origin
of the hot gas. We note that the cooling time ($\sim$~Hubble time) is
much longer than the enrichment time ($>$~1~Gyr) and the gas infall
time ($\sim$~300~Myr). It would also be consistent to argue that the
hot gas came from the ring at an early time when the ring abundance was
low, with the ring becoming enriched through its own subsequent star
formation. However, it also remains possible that the metal-poor hot
gas comes from the cold reservoir in the outer disk. Unfortunately,
there are no bright quasars in the vicinity of NGC~1291 to enable
measurement of the abundance of the outermost gas.

\section{Conclusions} 
\label{Conclusions}

The hot gas enrichment in ellipticals and early-type galaxies is
believed to be due mainly to type Ia SNe and stellar mass
loss. NGC~1291 has hot gas of low-metallicity at its centre, while
having a metal-rich stellar bulge. It is hard to reconcile the low
metallicity of the hot gas ($\sim 0.13 \times$ solar) with heating due to
stellar processes when the stars have metallicity $\sim 1.1 \times$ solar.
We have investigated the possibility that the hot gas comes from the
outer ring and is dynamically heated, by measuring the metallicities
of two HII regions, one in the ring and one closer to the centre. Both
regions show metallicities near solar, much higher than in the the
hot gas. Even allowing for a $\approx$~20\%  error in the X-ray
metallicity estimate due to the model dependence of the X-ray
abundances, the two metallicities cannot be brought together.

 However, when comparing the various timescales, there is time enough
for the gas in an initially metal-poor outer ring to become enriched
through the usual stellar evolution processes. Therefore, we cannot
exclude the possibility that the hot gas comes from an early
metal-poor outer ring and is dynamically heated. If the X-ray gas does
not come from the ring, as the abundances would at first sight
suggest, then it may still come from further out. Further evidence for
an external origin would be the presence of peculiar dust structure
and ionised gas in the lens and bar region. This infall from external
gas could be a common phenomenon, which we happen to detect here in
NGC~1291 only because its potential well is so deep. Following
the referee's comments and the results from this work, we have
started a campaign to do a systematic study on the relation between
external gas accretion and the metallicity of the X-ray emission in
early type galaxies.

\begin{acknowledgements}
      We thank Lisa Kewley for useful comments and for facilitating
      the use of her models which have made this work possible. We
      also thank the service observers at Paranal for the careful
      observations, and Edwin Valentijn  and Michiel Tempelaar who
      facilitated the use of the ESO-LV catalogue. We further thank
      the anonymous referee for the useful comments that helped us to
      improve this manuscript. This research has made use of the NASA/
      IPAC Infrared Science Archive, which is operated by the Jet
      Propulsion Laboratory, California Institute of Technology, under
      contract with the National Aeronautics and Space
      Administration. This work has also made use of the 2MASS Atlas
      Images.

\end{acknowledgements}
\bibliography{2687}
\bibliographystyle{natbib}
\end{document}